\documentclass[a4paper,twocolumn,showpacs,showpacs,superscriptaddress,prl,floatfix]{revtex4}
\usepackage{graphicx}          
\usepackage{amsmath}
\begin{document}
\title{Percolation threshold gold films on columnar coatings:\\\protect characterisation for SERS applications}

\author{Armandas Bal\v{c}ytis$^*$}
\affiliation{Centre for Micro-Photonics, Swinburne University of
Technology, John St., Hawthorn, VIC 3122, Australia}
\affiliation{State research institute Center for Physical Sciences
and Technology, Savanori\c{u} ave. 231, Vilnius, Lithuania,
LT-02300}

\author{Tomas Tolenis\footnote{$^*$ authors who contributed equally. }}
\affiliation{State research institute Center for Physical Sciences
and Technology, Savanori\c{u} ave. 231, Vilnius, Lithuania,
LT-02300}

\author{Xuewen Wang}
\affiliation{Centre for Micro-Photonics, Swinburne University of
Technology, John St., Hawthorn, VIC 3122, Australia}

\author{Gediminas Seniutinas}
\affiliation{Centre for Micro-Photonics, Swinburne University of
Technology, John St., Hawthorn, VIC 3122, Australia}

\author{Ramutis Drazdys}
\affiliation{State research institute Center for Physical Sciences
and Technology, Savanori\c{u} ave. 231, Vilnius, Lithuania,
LT-02300}

\author{Paul R. Stoddart}
\affiliation{Faculty of Science, Engineering and Technology,
Swinburne University of Technology, John St., Hawthorn, VIC 3122,
Australia}

\author{Saulius Juodkazis}
\affiliation{Centre for Micro-Photonics, Swinburne University of
Technology, John St., Hawthorn, VIC 3122, Australia}

\date{\today}
\begin{abstract}
Percolation of gold films of $\sim 15$~nm thickness was controlled
to achieve the largest openings during Au deposition. Gold was
evaporated on 300-nm-thick films of nanostructured porous and
columnar SiO$_2$, TiO$_2$ and MgF$_2$ which were deposited by
controlling the angle, rotation speed during film formation and
ambient pressure. The gold films were tested for SERS performance
using thiophenol reporter molecules which form a stable
self-assembled monolayer on gold. The phase retardation of these
SERS substrates was up to 5\% for wavelengths in the visible
spectral range, as measured by Stokes polarimetry. The SERS
intensity on gold percolation films can reach $\sim
10^3$~counts/(mW$\cdot$s) for tight focusing in air, while
back-side excitation through the substrate has shown the presence
of an additional SERS enhancement via the Fresnel near-field
mechanism.
\end{abstract}
\pacs{Keywords: 3D coatings, Raman sensors, surface enhanced Raman
scattering}
\maketitle
%

\section{Introduction}

It was demonstrated that optical near-field effects can be
utilised to augment surface enhanced Raman scattering (SERS)
intensity when light is traveling from the substrate (usually
glass) into air or solution from where analyte molecules can
attach to nano-islands of gold or other plasmonic
metal~\cite{13sr2335}. This SERS excitation geometry is usually
referred to as back-side whereas the standard retro-reflection
setup with laser beam irradiating a substrate with metal
nanoparticles from the air/solution side is the front-side.

\begin{figure*}[t]
\begin{center}
\includegraphics[width=16.50cm]{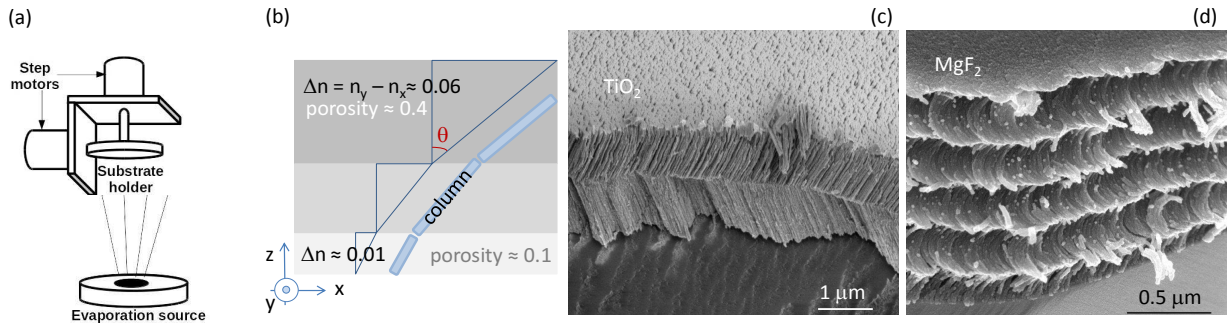}
\caption{Principle of glancing angle deposition. The orientation
of the substrate is controlled by two step motors. An evaporation
source is either an electron beam heated crucible with
Ti$_{3}$O$_{5}$ material or a resistively heated tungsten boat
with MgF$_{2}$ material. (b) Schematics of a columnar film grown
by the glancing angle deposition with controlled porosity and
birefringence $\Delta n$ typical for TiO$_2$~\cite{Lee}. Cross
sectional views of (c) TiO$_2$ chevron films made at two different
deposition angles and (d) MgF$_2$ chiral waveplate made by
glancing angle deposition with rotation.} \label{f-intro}
\end{center}
\end{figure*}

Metal films nanometers in thickness deposited on different
substrates have distinct formation morphologies defined by surface
energy and deposition conditions. Percolation of individual
islands takes place to form a film when thickness is typically
exceeds 10~nm. Since back-side Raman excitation can increase SERS
sensitivity by a factor of $\propto n^4$ where $n$ is the
refractive index of the substrate (in reference to air or
solution), it is of interest to explore properties of gold metal
films at the percolation threshold, where interconnected metal
islands produce the largest nano-groove openings. The possible
polarisation effects on light transmission through a random
pattern of nano-grooves, its coupling into surface plasmon
polaritons, as well as plasmon localization effects are all very
relevant for control of SERS performance. Recently, it was
demonstrated that patterns of nanoparticles forming a metamaterial
which can support a traveling surface wave are better suited for
SERS~\cite{Kabashin}. Also, randomness of a nanoparticle pattern
in terms of their size and separation is advantageous for the
spectrally broad-band performance and increased intensity in
SERS~\cite{14aom382,13oe13502}. The percolation metal films
studied here exhibit both randomness and surface
interconnectivity.

When $\sim \lambda/2$ waveplates are made out of dielectric
layers deposited on a flat surface, it should be possible to achieve
effective back-side irradiation field enhancement conditions using
standard front-side illumination. For the $\sim \lambda/2$
waveplate the phase shift of $2\pi$ occurs for light which
experiences reflection and a double pass in the film. By
engineering reflectivity of the substrate and waveplate material
with deposited gold films, SERS substrates can be realised with
the added benefit of back-side augmentation of the local E-field of light.
Such substrates can be straightforwardly used in upright Raman
microscopy tools, whilst making use of an increased sensitivity. Complex 3D
columnar coatings acting as waveplates have nano-rough surfaces
when grown by glanced angle deposition. This feature has potential
to be used for SERS after plasmonic metal coating.

Here, we report on the optical properties of gold films at the
percolation threshold deposited on nanostructured SiO$_2$, TiO$_2$
and MgF$_2$ coatings formed by glanced angle deposition and under
low vacuum conditions for their use as SERS substrates. Such
sensor surfaces are compatible with a practical fiber
platform~\cite{Gupta,Stoddart09}.

\section{Samples and procedures}


For experimental investigation six different samples were produced
using a thin film deposition machine VERA 1100 (VTD, Germany). All
the films were prepared on a silica substrate by first coating
dielectrics as the base layer and subsequently depositing gold on
top of them. The inner structures and materials used for
dielectric layers are summarized in Table~\ref{Table:Samples}.
Porosity in the dielectric films was due to intense gas-phase
collisions in low $P_{0}=10^{-1}$~Pa vacuum during transport from
the source to the substrates~\cite{Lakhtakia2005book}. The tilted
columnar thin film (CTF) structure was formed by using the the
glancing angle deposition~\cite{Messier1997} technique which is
used to make CTF and chiral films with controlled porosity and
birefringence (Fig.~\ref{f-intro}(a,b)). The substrates were
placed stationary above the evaporation source. By manipulating
the substrate with dual step motors exceptional variety of
achievable 3D film structures can be realized
(Fig.~\ref{f-intro}(c,d)). As part of this work the angle between
the vapour flux and the substrates was maintained at 60 or 70
degrees, depending on sample. For the \textbf{T60}\_TiO$_2$ and
\textbf{T70}\_TiO$_2$ samples the evaporation source was
Ti$_{3}$O$_{5}$ material heated using an electron beam. A 300~nm
thickness was coated at the evaporation rate of 2\AA/s. In order
to achieve complete oxidation, an additional oxygen flow of
20~sccm was introduced into the chamber. For the
\textbf{V}\_MgF$_2$ sample the evaporation source was MgF$_{2}$
material placed in a resistively heated tungsten boat. For the
production of vertical CTFs, the substrate was rotated about its
axis at 1 rev/s. The remainder of the substrate geometry was kept
the same. In CTF manufacturing cases the distance between the
substrate and the evaporation source was maintained at 30~cm to
ensure a collimated flux of vapour.

\begin{figure*}[t]
\begin{center}
\includegraphics[width=9.0cm]{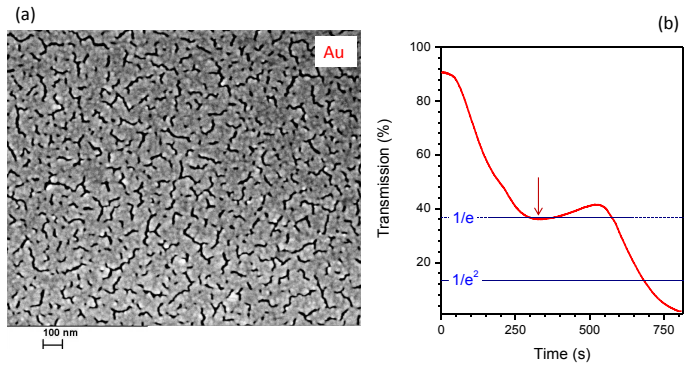}
\caption{(a) Gold film at the percolation threshold evaporated on
the  SiO$_2$ porous film (sample $\mathbf{P}$\_SiO$_2$). (b)
Transient of transmission monitored during gold deposition. Arrow
mark represents the condition of film shown in (a). } \label{f-trans}
\end{center}
\end{figure*}

\begin{table}[ht]
\center \caption{Summary of the materials used for dielectric
layers and their inner structure. The layers are porous (P),
columnar tilted (T) and vertical (V). CTF = columnar thin
film.}\label{Table:Samples} \vspace{0.5cm}
\begin{tabular}{|c|c|c|}\hline
Sample  & Structure& Material\\\hline\hline $\mathbf{P}$\_SiO$_2$
& $\mathbf{P}$orous & SiO$_2$   \\\hline $\mathbf{P}$\_MgF$_2$ &
$\mathbf{P}$orous & MgF$_2$
\\\hline $\mathbf{V}$\_MgF$_2$      & $\mathbf{V}$ertical CTF & MgF$_2$
\\\hline $\mathbf{T60}$\_TiO$_2$ & 60$^{\circ}$ $\mathbf{T}$ilted CTF & TiO$_2$ \\\hline $\mathbf{T70}$\_TiO$_2$ & 70$^{\circ}$ $\mathbf{T}$ilted CTF & TiO$_2$ \\\hline $\mathbf{V}$\_TiO$_2$
& $\mathbf{V}$ertical CTF & TiO$_2$  \\\hline
 \end{tabular}
\end{table}

The top gold layer was evaporated in a separate process at the rate of 0.5\AA/s for all
the samples simultaneously. A thickness between 10 and 20~nm was
formed using a resistive evaporator. A flat witness glass was used to
control the process by monitoring the transmission at the
wavelength of 600~nm. The evaporation process was stopped when the
transmission reached the first minimum due to
the percolation effect (Fig.~\ref{f-trans}).

\begin{figure}[tb]
\begin{center}
\includegraphics[width=8.5cm]{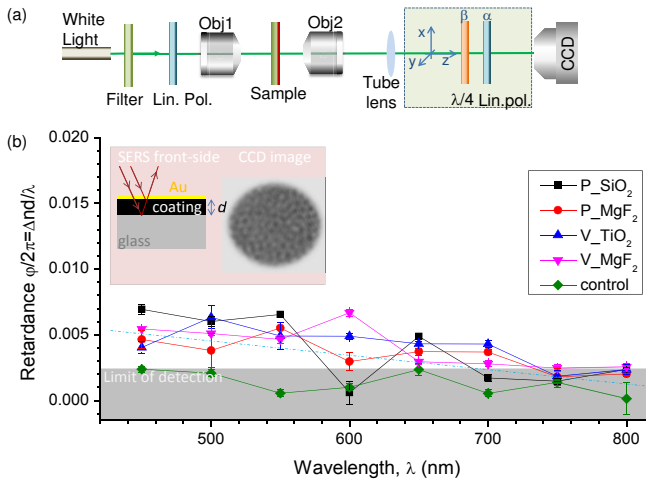}
\caption{(a) Schematic of the setup used to measure Stokes parameters $S(I,Q,U,V)$ and
retardance $\varphi = \mathrm{atan}(V/U)$. (b) Retardance,
$\varphi$, spectra presented in wavelength fractions $\varphi/2\pi = \Delta
n d/\lambda$ for different samples with porous and columnar films
(Table~\ref{Table:Samples}); 1-mm-thick silica was used as a control sample. Inset shows schematics of the front-side SERS
measurements and a CCD image of a $\sim 50~\mu$m-diameter area on
the sample which was integrated in order to determine the Stokes
parameters. Error bars are $\pm 30\%$. The dashed line is
prediction of a linear dependence $\varphi\propto d/\lambda$. }
\label{f-retar}
\end{center}
\end{figure}

\begin{figure*}[tb]
\begin{center}
\includegraphics[width=16cm]{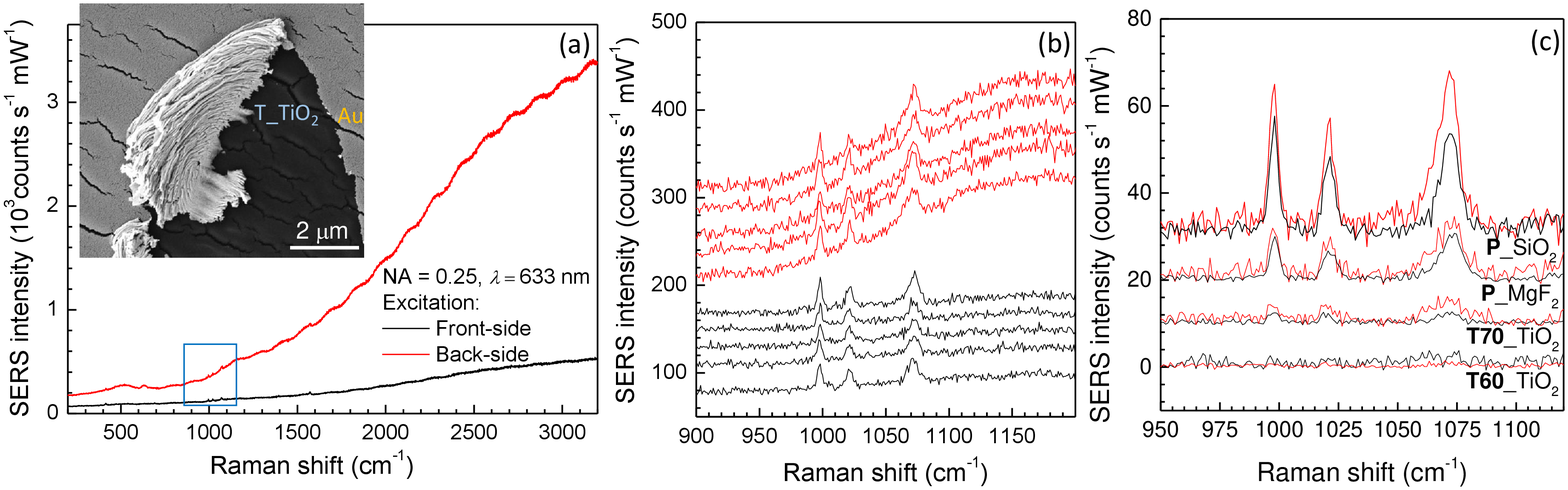}
\caption{ Raman scattering spectra of thiophenol measured on
$\sim 15$~nm thick Au via front-side and back-side acquisition at
several locations. (a) Broad-sweep Raman spectra obtained on sample $\mathbf{P}$\_SiO$_2$ and exhibiting increased background scattering for the $\sim4$~mm thickness glass
substrate. Weak thiophenol signature is shown in the
region-of-interest box; Inset
shows a typical SEM image of surface used for SERS: Au on the CTF
film. (b) Close-up view of thiophenol spectral lines measured on sample
$\mathbf{P}$\_SiO$_2$ in
top-side and back-side acquisition cases. (c) Comparison of background subtracted
Raman signal obtained on different dielectric coatings underlying
the thin Au film at the percolation threshold. Objective lens was $NA
= 0.25$, wavelength of excitation $\lambda = 633$~nm, acquisition
time 5~s. See Table~\ref{Table:Samples} for sample nomenclature.}
\label{f-sers}
\end{center}
\end{figure*}

Polariscopy measurements were used to characterise the
polarisation changes of light propagating through the columnar film and Au
coating. The intensity of the transmitted light, $I_T$, was
measured using a $\lambda/4$ waveplate and an analyser at different
orientational angles $\beta$ and $\alpha$, respectively, according
to~\cite{Berry1977}:
\begin{equation}\label{e1}
\begin{aligned}
I_T(\alpha,\beta) = &1/2[I+(Q\cos(2\beta) +
U\sin(2\beta))\cos(2(\alpha-\beta))\\
& + V\sin(2(\alpha-\beta))],
\end{aligned}
\end{equation}
\noindent where $\alpha$ and $\beta$ are the rotation angles of
the fast axis of the polarizer and the waveplate with respect to
the $x$-axis (horizontal).

The four Stokes parameters, $S(I,Q,U,V)$, that define an arbitrary
state of light polarisation, are determined by separate
measurements with the waveplate and analyser at fixed angles of
$\alpha$ and $\beta$, given by:
\begin{equation}\label{e2}
\begin{aligned}
I &= I_T(0,\pi/4) + I_T(0,-\pi/4),\\ Q &= 2I_T(0,0) -I,\\ U &=
2I_T(\pi/4,\pi/4) - I,\\ V &= 2I_T(0,-\pi/4) - I.
\end{aligned}
\end{equation}
\noindent It is usual to present Stokes parameters as the
Poincar\'{e} sphere with the retardation given by $\varphi =
\mathrm{atan}(V/U)$.

Filters of 10~nm spectral bandwidth were used to select different
central wavelengths from a white-light super-continuum laser
(SuperK Compact, NKT Photonics) in the range from 400 to 800~nm
and focused to a $5 - 7\mu$m spot on the sample. The
incident linearly polarised light was set at 45~degree polarization using a
Glan-Taylor polarizer and was focused onto a region of the sample via an objective lens of numerical aperture $NA=0.26$ ($10 \times$, Mitutoyo). After the sample, an $NA=0.5$
objective lens ($100 \times$, Mitutoyo) was used to collect and
collimate the transmitted light for polarisation analysis. The
sample and the two objectives were mounted on 2D positioning stages.
Polarisation analysis was carried out after the second objective
with an achromatic $\lambda/4$-waveplate and a second Glan-Taylor
polarizer. Phase retardation, $\varphi$, of the $45^\circ$ linear polarized beam
(i.e. the phase delay between the $s-$ and $p-$polarized light)
was measured as $\varphi=2\pi\Delta nd/\lambda$, where $\Delta n$
is the birefringence, $d$ is the thickness of the structure (a film with
Au coating), and $\lambda$ is the wavelength of light.  A tube lens and a $1024\times 768$~pixel CCD were used for imaging and transmission measurements. Typical lateral
resolution was $5 - 7~\mu$m. The setup was tested by measuring
retardance of a commercial $\lambda/2$-plate.

Raman scattering spectra were acquired using an \mbox{InVia}
Streamline Raman microscope (Renishaw, UK) under $\lambda =
633$~nm continuous wave laser excitation at $P = 2.94$~mW  total
power, focused with an objective lens of numerical aperture $NA =
0.25$. The Au layer was functionalized by immersing the substrates
for 30~min in a 10~mM ethanolic thiophenol solution. Substrates
where subsequently rinsed in ethanol, to remove thiophenol
molecules not bound to Au, and blow-dried with N$_{2}$. Near-field
enhancement effects were probed by performing Raman spectra
acquisitions, both, with laser incidence from the thin-film
deposition side (front-side) and by focusing on the Au layer
through the $~ 4$~mm-thick glass substrate (back-side).

\section{Results and discussion}

For SERS functionality nano-particles and nano-gaps are essential
for light field enhancement at nanoscale - the generation of
``hot-spots''. Gold films at the percolation threshold have a
random pattern of openings, hence, nano-gaps which can be used as
hot-spots in SERS. Figure~\ref{f-trans}(a) shows a typical Au film
coated over a SiO$_2$ porous layer (sample $\mathbf{P}$\_SiO$_2$).
This surface morphology corresponds to the local minimum in the
transmission observed during deposition (Fig.~\ref{f-trans}(b))
and is also observed when TiO$_2$ and MgF$_2$ underlying
dielectric coatings are used. Thickness of the gold film was
10-20~nm as could be estimated from SEM images. Gold was
evaporated directly onto the porous or CTFs and was loosely
attached to the substrate. A local transmission minimum occurs
when transmission, $T$, is approximately at a $1/e$-level (for the
normalised transmission). The refractive index of gold at $\lambda
= 600$~nm is $n+ik = 2.42 +i2.9152$, hence, the transmission
through a thickness of $x = 16.5$~nm is  $T=\exp(-4\pi kx/\lambda)
\simeq 1/e$. For longer gold deposition durations there is at
first a slight increase in transmission, most probably due to an
augmented forward scattering through the openings as result of
Fresnel reflection, before $T$ starts to continuously decrease
once the openings are closed and the gold film grows in thickness.

Columnar coatings deposited at large glancing angles exhibit
nanoscale surface roughness and by applying a thin Au coating they were
prepared for SERS. Polariscopy measurements were carried out to
characterise the retardance of the coating with Au film at the
percolation threshold. Figure~\ref{f-retar} shows that an
effective retardance measured in transmission was only 0.2-0.5\%
of the corresponding wavelength over the entire visible spectrum.
This is a very small number and the porous and columnar films
served only as nano-rough substrates for SERS. It is noteworthy,
that coatings corresponding to $\lambda/2$ retardance would realise
phase matching for constructive interference on the Au surface
(inset in Fig.~\ref{f-retar}) and can be used to augment SERS
intensity (planned as a next step of this study). The light
reflected from the \mbox{glass-coating} boundary (inset in
Fig.~\ref{f-retar}) can be considered as a back-side SERS
excitation component, since it is propagating from the high to the low
refractive index region, hence, the Fresnel enhancement takes
place~\cite{13sr2335}. Thickness of the coating, $d$, was measured
from SEM  cross sectional images and was $d \simeq 300\pm 10$~nm,
which defines the birefringence $\Delta n \simeq 6\times 10^{-3}$
for the measured $\varphi/2\pi = 0.003$ at $\lambda = 633$~nm used
in SERS. Intriguingly, a very small birefringence was observed in
the case of porous and vertical columnar structures
(Fig.~\ref{f-retar}) for which isotropic, hence non-birefringent, film performance is expected. However, film deposition is a complex
process governed by local temperature and material delivery
peculiarities, hence, local symmetry braking and anisotropy can
be expected and related to the material flux. For example, surface
of the vertical columnar film (Fig.~\ref{f-intro}(d)) clearly
shows an anisotropy of nano-grain orientation on the top of the
film. Thicker dielectric films will be made for experimental
measurements of birefringence and will be related to the cross
sectional morphology and porosity in a following study.

Low SERS signals were observed from thiophenol reporter
molecule coated samples (Fig. ~\ref{f-sers}). The back-side
irradiation mode showed a strong fluorescence background
(Fig.~\ref{f-sers}(a)). The region-of-interest for thiophenol
peaks on porous silica $\mathbf{P}$\_SiO$_2$ for the front- and
back-side SERS collection modes is shown in
(Fig.~\ref{f-sers}(b)). There was a recognisable difference
between SERS intensity for the two acquisition modes. The SERS
intensity is normalised only to the transmission through the
microscope and a front-side reflection is not taken into account
for the data shown in Fig.~\ref{f-sers} for the back-side mode
(reflectivity of the air-surface is $R = (n-1)^2/(n+1)^2$ for
intensity). A slightly larger SERS signal by a factor of 1.2-1.6
was observed for the back-side irradiation. The reason of the
enhancement is due to the local E-field amplitudes being different
in the surface for the air-substrate and substrate-air
cases~\cite{13sr2335}. For example, in the case of glass substrate
for a back-side mode the amplitude of the transmitted light
depends on the refractive index ratio between media of incidence
and transmission ($n_i = 1.5$ vs $n_t = 1$) as $t = 2n_i/(n_i
+n_t) \equiv 1.2$. Since SERS enhancement factor depends on the
local E-field intensity at the wavelength of excitation and
scattering - the electromagnetic enhancement factor, $M_{Enh}$ -
one can expect an augmented SERS intensity by a factor of $M_{Enh} =
I_{exc}^2\times I_{Raman}^2 \simeq t^4 \simeq 2.1$. The strongest
back-side enhancement was observed for a silica porous film which
has a slightly larger refractive index than porous MgF$_2$: for a
solid film it would correspond to 1.4 vs 1.2. Interestingly,
tilted column TiO$_2$ films (deposition at 60 and 70 degrees)
showed almost identical SERS intensity for both modes of excitation,
most likely due to the negligible difference in retardance in both cases over the 300 nm thickness. However, for the CTF structures,
there is a possibility to control the local light field intensity
on the interface where SERS is measured by the phase control through film thickness and
exploiting the zero phase shift for light traversing form the higher to lower refractive index region ($n_t<n_i$).

Earlier studies showed that SERS intensity is proportional to the
thickness of gold~\cite{12pa5,15n18299} for measurements in
reflection (front-side mode). Formation of hot-spots in SERS
occurs when 3D morphology of the surface develops in thicker gold
films. If gold is coated over existing 3D nano-textured surfaces
the hot-spots are formed in the crevices as demonstrated for well
controlled 3D surface patterns~\cite{15ami27661}. In this regard
the percolation nano-gaps are not as directly applicable for SERS
since they are mostly planar and formed on a flat substrate. SERS
is typically measured on 100-nm-thick Au coatings and up to 6-8
fold decrease in SERS intensity is expected on 16~nm gold films.
SERS intensity is also scales linearly with the solid collection
angle and a low numerical aperture $NA = 0.25$ (cone angle $\psi =
14.5^\circ$) used in back-side SERS measurements contributes to
the lower signal~\cite{12ps283}. The observed SERS rate of
30~counts/(mW$\cdot$s) would be 3 times stronger for the popular
$NA = 0.75$ (cone angle $\psi = 48.5^\circ$) acquisition mode.
This shows that the layer system of nanostructured dielectric and
thin gold films used in this study can perform at the level of
$10^3$~counts/(mW$\cdot$s) for the commonly used $NA = 0.75$ lens
which is typical for many reported SERS substrates. Further SERS
enhancement is achievable with formation of deep nano-grooves
usually obtained on nano-rough substrates using thick ($d >
100$~nm) metal coatings.

\section{Conclusions and Outlook}

It has been demonstrated that thin $\sim 15$~nm gold coatings at
the percolation threshold can be used as SERS substrates.
Re-scaled for acquisition with a typical $NA = 0.75$ objective
lens, the SERS intensity could reach $10^3$~counts/(mW$\cdot$s),
which is typical in the SERS field. The presence of the back-side
SERS intensity augmentation is confirmed for percolation films for
the first time.

Future work is required for fabricating SERS substrates which are
optimised for maximum E-field enhancement factors on the surface
via the use of birefringent CTF $\lambda/2$ plates and percolation
coatings. Phase control via the thickness of the phase plate could
bring a new means of SERS control~\cite{Grigorenko} and can be
implemented to improve the signal-to-noise ratio using SERS
signals from two different CTF regions. Such control would
inherently require a percolation property of films for light
transmission. Porosity and gas permeability of CTF could also find
use in practical SERS applications.

\subsection{Acknowledgements}

SJ is grateful for partial support via the Australian Research
Council Discovery project DP130101205 and a sensor technology
transfer project with Workshop of Photonics Ltd.

\small

\begin{thebibliography}{10}

\bibitem{13sr2335}
S.~Jayawardhana, L.~Rosa, S.~Juodkazis, and P.~R. Stoddart,
``Additional
  enhancement of electric field in surface-enhanced {Raman} scattering due to
  {Fresnel} mechanism,'' {\em Sci. Rep.}~{\bf 3}, p.~2335, 2013.

\bibitem{Lee}
G.~J. Lee, Y.~P. Lee, B.~Y. Jung, S.~G. Jung, C.~K. Hwangbo, J.~H.
Kim, and
  C.~K. Yoon, ``Microstructural and nonlinear optical properties of thin silver
  films near the optical percolation threshold,'' {\em J. Korean Phys.
  Soc.}~{\bf 51}(4), pp.~1555 -- 1559, 2007.

\bibitem{Kabashin}
A.~V. Kabashin, P.~Evans, S.~Pastkovsky, W.~Hendren, G.~A. Wurtz,
R.~Atkinson,
  R.~Pollard, V.~A. Podolskiy, and A.~V. Zayats, ``Plasmonic nanorod
  metamaterials for biosensing,'' {\em Nature Materials}~{\bf 8}, pp.~867 --
  871, 2009.

\bibitem{14aom382}
Y.~Nishijima, Y.~Hashimoto, L.~Rosa, J.~B. Khurgin, and
S.~Juodkazis, ``Scaling
  rules of {SERS} intensity,'' {\em Adv. Opt. Mat.}~{\bf 2}(4), pp.~382--388,
  2014.

\bibitem{13oe13502}
Y.~Nishijima, J.~B. Khurgin, L.~Rosa, H.~Fujiwara, and
S.~Juodkazis,
  ``Randomization of gold nano-brick arrays: a tool for {SERS} enhancement,''
  {\em Opt. Express}~{\bf 21}(11), pp.~13502--13514, 2013.

\bibitem{Gupta}
A.~K. Sharma, R.~Jha, and B.~D. Gupta, ``Fiber-optic sensors based
on surface
  plasmon resonance: a comprehensive review,'' {\em Sensors Journal
  {IEEE}}~{\bf 7}(8), pp.~1118--1129, 2007.

\bibitem{Stoddart09}
P.~R. Stoddart and D.~J. White, ``Optical fibre {SERS} sensors,''
{\em Anal.
  Bioanal. Chem.}~{\bf 394}(7), pp.~1761 -- 1774, 2009.

\bibitem{Lakhtakia2005book}
A.~Lakhtakia and R.~Messier, {\em Sculptured thin films:
Nanoengineered
  Morphology and Optics}, SPIE Press, Bellingham, Washington USA, 2005.

\bibitem{Messier1997}
R.~Messier, T.~Gehrke, C.~Frankel, V.~C. Venugopal, W.~Otano, and
A.~Lakhtakia,
  ``Engineered sculptured nematic thin films,'' {\em J. Vac. Sci. Technol.
  A}~{\bf 15}(4), pp.~2148--2152, 1997.

\bibitem{Berry1977}
H.~G. Berry, G.~Gabrielse, and A.~E. Livingston, ``Measurement of
the {Stokes}
  parameters of light,'' {\em Applied Optics}~{\bf 16}(12), pp.~3200 -- 3205,
  1977.

\bibitem{12pa5}
R.~Buividas, P.~R. Stoddart, and S.~Juodkazis, ``Laser fabricated
ripple
  substrates for surface-enahnced {Raman} scattering,'' {\em Annalen der
  Physik}~{\bf 524}(11), pp.~L5 -- L10, 2012.

\bibitem{15n18299}
A.~Bal\v{c}ytis, M.~Ryu, G.~Seniutinas, J.~Juodkazyt\.{e},
B.~C.~C. Cowie,
  P.~R. Stoddart, J.~Morikawa, and S.~Juodkazis, ``Black-{CuO}:
  Surface-enhanced {Raman} scattering and infrared properties,'' {\em
  Nanoscale}~{\bf 7}(43), pp.~18299--18304, 2015.

\bibitem{15ami27661}
S.~Dinda, V.~Suresha, P.~Thoniyot, A.~Bal\v{c}ytis, S.~Juodkazis,
and
  S.~Krishnamoorthy, ``Engineering {3D} nanoplasmonic assemblies for high
  performance in spectroscopic sensing,'' {\em ACS Appl. Mater. Interf.}~{\bf
  7}(50), pp.~27661 -- 27666, 2015.

\bibitem{12ps283}
S.~Jayawardhana, L.~Rosa, R.~Buividas, P.~R. Stoddart, and
S.~Juodkazis,
  ``Light enhancement in surface-enhanced {Raman} scattering at oblique
  incidence,'' {\em Photonic Sensors}~{\bf 2}(3), pp.~283--288, 2012.

\bibitem{Grigorenko}
A.~N. Grigorenko, P.~I. Nikitin, and A.~V. Kabashin, ``Phase jumps
and
  interferometric surface plasmon resonance imaging,'' {\em Appl. Phys.
  Lett.}~{\bf 75}(25), pp.~3917--3919, 1999.

\end{thebibliography}

\end{document}